\documentclass[proof]{WileyASNA-v1}

\articletype{Article Type}%

\received{22 Sept 2020}
\revised{10 Oct 2020}
\accepted{19 Oct 2020}

\begin{document}

\title{Energy Conditions in Non-minimally Coupled $f(R,T)$ Gravity}

\author[1]{P.K. Sahoo*}

\author[1]{Sanjay Mandal}

\author[1]{Simran Arora}

\authormark{P.K. Sahoo \textsc{et al}}

\address[1]{\orgdiv{Department of Mathematics}, \orgname{Birla Institute of Technology and
Science-Pilani}, \orgaddress{\state{Hyderabad Campus, Hyderabad-500078}, \country{India}}}

\corres{*Corresponding author name,  \email{pksahoo@hyderabad.bits-pilani.ac.in}}

\abstract{In today's scenario, going beyond Einstein's theory of gravity leads us to some more complete and modified gravity theories. One of them is the $f(R,T)$ gravity in which $ R $ is the Ricci scalar, and $ T $ is the trace of the energy-momentum tensor. Using a well-motivated linear $f(R,T)$ gravity model with a single parameter, we studied the strong energy condition (SEC), the weak energy condition (WEC), the null energy condition (NEC), and the dominant energy condition (DEC)  under the simplest non-minimal matter geometry coupling with a perfect fluid distribution. The model parameter is constrained by energy conditions and a single parameter proposed equation of state (EoS), resulting in the compatibility of the $f(R,T)$ models with the accelerated expansion of the universe. It is seen that the EoS parameter illustrate the quintessence phase in a dominated accelerated phase, pinpoint to the cosmological constant yields as a prediction the phantom era. Also, the present values of the cosmological constant and the acceleration of the universe are used to check the viability of our linear $f(R,T)$ model of gravity. It is observed that the positive behavior of DEC and WEC  indicates the validation of the model. In contrast, SEC is violating the condition resulting in the accelerated expansion of the universe.}


\maketitle

\section{Introduction}\label{sec1}

Verification of the universe late-time acceleration has led to comprehensive studies into its interpretation. Observational data on type Ia supernovae revealed an acceleration of the current universe \citep{Riess1998,Riess1999,Perlmutter1999}. This phenomenon can either be overruled by introducing a new kind of universal fluid known as dark energy or proposing modification and variations of the Einstein theory of gravity. This is done by modifying the left-hand side of the Einstein equation, i.e., the spacetime's geometry. Modified theories of gravity are the geometrical generalizations of General Relativity (GR) in which the Einstein-Hilbert action is modified by replacing the Ricci scalar R with a more general function along with matter-geometry coupling. Some widely known and used modified theories are $f(R)$ gravity \citep{Nojiri2003}, $f(T)$ gravity \citep{Femaro2007}, $f(G)$ gravity \citep{Nojiri2008}, $f(R,T)$ gravity \citep{Harko2011}, $f(Q,T)$ gravity \citep{Yixin2019} etc. In the last few years, $f(R,T)$ gravity gained much attention where $R$ is the Ricci scalar, and $T$ is the trace of the energy-momentum tensor.

A lot of research on various aspects of $f(R,T)$ gravity such as thermodynamics \citep{Sharif2012}, scalar pertubations \citep{Alvarenga2013}, wormhole solutions  \citep{Moreas2017} has been discussed in literature. Apart from these effective experiments, any modified theory of gravity should follow an important role in the energy conditions. These conditions define the attractive existence and spacetime casual and geodesic structure in modified gravity.

The energy conditions are the necessary conditions for a good understanding of the singularity theorem, such as black-hole thermodynamics. The well known Raychaudhuri equations \citep{Ray1955} play a role in describing the attractive nature of gravity and positive energy density. The four basic conditions are the null energy condition (NEC), weak energy condition (WEC), dominant energy condition (DEC), and strong energy condition (SEC). All the conditions are derived from the Raychaudhuri equation, which helps us analyze the entire spacetime structure without precise solutions to Einstein's equations playing a vital role in understanding cosmological gravitational interactions. The null energy condition (NEC) discusses the second law of black-hole thermodynamics, although its violation corresponds to the universe's big rip singularity \citep{Carroll2004}. On the other strong energy condition (SEC) is useful for studying the Hawking-Penrose theorem of singularity \citep{Hawking1973}. SEC is also good at describing the repulsive/attractive nature of gravity under modified theories of gravity. The violation of SEC implies the observed accelerated expansion of the universe. Many works on energy conditions under modified theories are presented in the literature. K. Bamba et al. \citep{Bamba2017} studied the energy conditions in $f(G)$ gravity. They considered some realistic $f(G)$ forms that could be used in the late-time cosmic acceleration epochs to cure finite-time future singularities. S. Capozziello et al. present the role of energy conditions in $f(R)$ cosmology \citep{Capo2018}. K. Atazadeh et al. \citep{Atazadeh2014} also considered two forms of $f(R,G)$ gravity accounting the stability of the cosmological solutions using various energy conditions. M. Zubair et al. \citep{Zubair2015} worked on energy conditions in $f(T)$ gravity with non-minimal torsion-matter coupling etc.

As it is known, modified gravity theories can be studied in two separate groups based on a coupling between matter and geometry, i.e., minimal coupling and non-minimal coupling. In general, the minimal coupling is weak, whereas non-minimal coupling is a strong coupling. Several works investigate the cosmological implications of a coupling of non-minimal matter-geometry in $f(R,T)$.  P.H.R.S. Moraes and P.K. Sahoo \citep{Pedro2017} studied the most straightforward non-minimal matter-geometry coupling in the $f(R,T)$ cosmology. L.K. Sharma et al. \citep{Lokesh2018} also worked on the existence of non-minimal matter-geometry coupling govern by power-law in Bianchi I spacetime. Recently, P.K. Sahoo and S. Bhattacharjee \citep{Sahoo2020} investigated gravitational Baryogenesis in non-minimal coupled $f(R,T)$ gravity. Furthermore, most of the above studies are done by the prior assumptions of scale factor or followed by the parametrization techniques. However, working on these types of methods induce some unclear doubts on constraining the model parameters for the current value of the universe's age and cosmological quantities. To overcome this issue, we have been adopted the cosmography to study the present accelerated scenario of the universe in our work.

In this work, we considered non-minimally coupled $f(R,T)$ gravity i.e. $f(R,T)= R+\alpha R T$ where $\alpha$ is the model parameter, to study various energy conditions. Also, to check the viability of $f(R,T)$ gravity theory, we used the present values of the cosmological parameters $q_{0}$ and $H_{0}$. The equation of state parameter $\omega$ provides an acceptable candidate for comparing our models with $\Lambda$CDM. The recent results from Planck Collaboration \citep{Planck} and the $\Lambda$CDM model indicate that the equation of state parameter $\omega\simeq -1$. This behavior corresponds to the Universe's negative pressure framework, which specifies the current accelerated phase.

The literature includes several works with certain linear forms of $f(R, T)$ gravity \citep{Psahoo2020, Sahoo2016, Harko2011}. We have considered the linear form of $f(R, T)$ gravity in the last second section, which demonstrates some consistency with $\Lambda$CDM and non-linear form. Using restricted values of cosmological parameters, energy conditions are examined, which is different from work done earlier.

The paper is presented and organized as follows: In section \ref{sec2}, we briefly describe the formulation of field equations in $f(R,T)$ gravity. In the next section, \ref{sec3}, we discuss the solutions of the field equations. The various energy conditions are studied in section \ref{sec4}. Section \ref{sec4} presents the comparison with the $\Lambda$CDM model along with the equation of state parameter. Also, the compatibility of model in case of linear form is presented in section \ref{sec5}. Further discussions and conclusion are presented in section \ref{sec6}.

\section{Basic equations of $f(R,T)$ gravity}\label{sec2}

The total action in the $f(R,T)$ theory of gravity reads \citep{Harko2011}
\begin{equation}\label{1}
S=\frac{1}{16\pi}\int d^{4}x\sqrt{-g}f(R,T)+\int d^{4}x\sqrt{-g}L_m,
\end{equation}
with $g$ being the metric determinant and $L_m$ be the matter Lagrangian density, and the stress-energy tensor of matter is defined as
\begin{equation}
\label{1a}
T_{\mu\nu}=-\frac{2}{\sqrt{-g}}\frac{\delta \left(\sqrt{-g}L_m\right)}{\delta g^{\mu\nu}}.
\end{equation}

By varying this action \eqref{1} with respect to the metric yields
\begin{eqnarray}\label{2}
\Big[f_1'(R) & + & f_2'(R)f_3(T)\Bigr]R_{\mu\nu}-\frac{1}{2}f_1(R)g_{\mu\nu} \\ \nonumber
& +  &
\Bigl(g_{\mu\nu}\Box  -  \nabla_\mu\nabla_\nu \Bigr)\Bigl[f_1'(R)+f_2'(R)f_3(T)\Bigr] \\ \nonumber
 =  \Bigl[8\pi & + & f_2(R)f_3'(T)\Bigr]T_{\mu\nu}+ f_2(R)\left[f_3'(T)p+\frac{1}{2}f_3(T)\right]g_{\mu\nu},
\end{eqnarray}
for which it was assumed $f(R,T)=f_1(R)+f_2(R)f_3(T)$ and primes denote derivatives with respect to the argument.

Now, we will take $f_1(R)=f_2(R)=R$ and $f_3(T)=\alpha T$, with $\alpha$ a constant. This is the simplest non-trivial functional form of the function $f(R,T)$ which involves matter-geometry coupling within the $f(R,T)$ formalism. Moreover, it benefits from the fact that General Relativity is retrieved when $\alpha=0$.
Eq.\eqref{2} now reads
\begin{equation}\label{3}
G_{\mu\nu}=8\pi T_{\mu\nu},
\end{equation}
with  $T_{\mu\nu}$ the usual matter energy-momentum tensor. 

\section{The $f(R,T)=R+\alpha RT$ cosmology} \label{sec3}

For a flat Friedmann-Robertson-Walker universe with scale factor $a(t)$ and Hubble parameter $H=\dot{a}/a$, the expressions for energy density and pressure from Eq. \eqref{2} as obtained in \citep{Pedro2017} reads as
\begin{equation}\label{4}
\rho\!=\!\frac{H^2 \! \left[8\pi-27\alpha\left(\dot{H}+2H^2\right)\right]\!+\!7\alpha(2\dot{H}+3H^2)\left(\dot{H}+2H^2\right)}{\frac{64\pi^2}{3}-96\pi \alpha \left(\dot{H}+2H^2\right)+18\alpha^2 \left(\dot{H}+2H^2\right)^2},
\end{equation}

\begin{equation}\label{5}
p\!=\!-\frac{9\alpha H^2 \! \left(\dot{H}+2H^2\right)\!+\!\left(2\dot{H}\!+\!3H^2\right) \! \left[\frac{8\pi}{3}-3\alpha\left(\dot{H}+2H^2\right)\right]}{\frac{64\pi^2}{3}-96\pi \alpha \left(\dot{H}+2H^2\right)+18\alpha^2 \left(\dot{H}+2H^2\right)^2}.
\end{equation}

The study of cosmological scenarios is done by researchers using various methods. Like the parametrization technique, this method is a good method to display the universe's profiles through mathematical modeling.  Nevertheless, some doubts about constraining the model parameters against the observational data for present-day values of cosmological quantities are unclear. Keeping this in mind, one can adopt the cosmography, which can overcome the previous issues. The cosmographic parameters can be derived from the expansion of scale factor concerning cosmic time. The Taylor series expansion of the scale factor up to its fifth-order is given by
\begin{eqnarray}\label{5a}
 a(t) & = & a(t_0)\Bigl\{H_0(t-t_0)-\frac{q_0}{2}H_0^2(t-t_0)^2+\frac{j_0}{3!}H_0^3(t-t_0)^3 \nonumber\\
&+ & \frac{s_0}{4!}H_0^4(t-t_0)^4+\frac{l_0}{5!}H_0^5(t-t_0)^5+\mathcal{O}[(t-t_0)^6]\Bigr\},
\end{eqnarray}
where the cosmological parameters such as Hubble parameter, deceleration parameter, jerk, snap, and lerk can be written as
\begin{eqnarray}\label{6}
H = \frac{1}{a}\frac{da}{dt} ; \, \, \,  \, \, \,  
q = -\frac{1}{aH^2}\frac{d^{2}a}{dt^2} ; \nonumber \\ 
j = \frac{1}{aH^3}\frac{d^{3}a}{dt^3} ; \, \, \, \, \, \,  
 s = \frac{1}{aH^4}\frac{d^{4}a}{dt^4} ; \nonumber \\
  l =\frac{1}{aH^5}\frac{d^{5}a}{dt^5}.
\end{eqnarray}

Also, we can customize the above terms as follows
\begin{eqnarray}\label{7}
\dot{H} & = & -H^2(1+q) \, ; \\
\label{8}
\ddot{H}& = & H^3(j+3q+2) \, ; \\ \label{8a}
\dddot{H} & = & H^4[s-4j-3q(q+4)-6] \, .
\end{eqnarray}
Now, using Equations \eqref{7}, \eqref{8} in \eqref{4} and \eqref{5}, we got the following expressions
\begin{equation}\label{10}
\rho=\frac{3 \alpha  H^4 (q-1) (7 q+10)+12 \pi  H^2}{27 \alpha ^2 H^4 (q-1)^2+144 \pi  \alpha  H^2 (q-1)+32 \pi ^2} \, ; 
\end{equation}
\begin{equation}\label{11}
p=\frac{H^2 \left(9 \alpha  H^2 \left(q^2-1\right)+\pi  (8 q-4)\right)}{27 \alpha ^2 H^4 (q-1)^2+144 \pi  \alpha  H^2 (q-1)+32 \pi ^2} \, . 
\end{equation}

\section{Energy Conditions}\label{sec4}

The energy conditions play a vital role in understanding the geodesics of the Universe. These conditions are derived from the well-known Raychaudhuri equations. For a congruence of time-like and null-like geodesics, the Raychaudhuri equations are given in the following forms \citep{Ehlers2006,Nojiri2007}
\begin{equation}\label{12}
\frac{d\theta}{d\tau}=-\frac{1}{3}\theta^2-\sigma_{\mu\nu}\sigma^{\mu\nu}+\omega_{\mu\nu}\omega^{\mu\nu}-R_{\mu\nu}u^{\mu}u^{\nu}\,,
\end{equation}
and 
\begin{equation}\label{13}
\frac{d\theta}{d\tau}=-\frac{1}{2}\theta^2-\sigma_{\mu\nu}\sigma^{\mu\nu}+\omega_{\mu\nu}\omega^{\mu\nu}-R_{\mu\nu}n^{\mu}n^{\nu}\,,
\end{equation}
where $\theta$ is the expansion factor, $n^{\mu}$ is the null vector, and $\sigma^{\mu\nu}$ and $\omega_{\mu\nu}$ are, respectively, the shear and the rotation associated with the vector field $u^{\mu}$. For attractive gravity, equations \eqref{16}, and \eqref{17} satisfy the following conditions
\begin{equation}\label{14}
R_{\mu\nu}u^{\mu}u^{\nu}\geq0 \, ; \,\,\,\,\,
 R_{\mu\nu}n^{\mu}n^{\nu}\geq0\,.
\end{equation}
For perfect fluid matter distribution, the above conditions read the following
\begin{itemize}
\item Strong energy conditions (SEC):  $\rho+3p\geq 0$;

\item Weak energy conditions (WEC):  $\rho\geq 0, \rho+p\geq 0$;

\item Null energy condition (NEC):  $\rho+p\geq 0$;

\item Dominant energy conditions (DEC): $\rho\geq 0, |p|\leq \rho$.
\end{itemize}

Planck collaborations have recently improved the calculation of cosmological parameters relative to the findings of 2015, leading to substantial improvements in the accuracy of correlated parameters. The work assumed the cosmology of the base $\Lambda$CDM and observed the parameters of the late universe. So, it is found that the value of $H_0$ is $(67.4 \pm 0.5)$ km s$^{-1}$ Mpc$^{-1}$ \citep{Planck}. Capozziello et al. also observed the cosmographic parameters $q_0, s_0, j_0$ values \citep{Capozziello2019}. The cosmographic techniques are implemented to derive cosmological consistent models of modified gravity. Further the value of $q_0$ is observed to be $-0.503$.

To check the viability of $f(R,T)$ gravity theory we use the present values of the cosmological parameters as $H_0=67.9$ and $q_0=-0.503$. The above conditions reads as follows
\begin{equation}\label{15}
\rho+3p\!=\!\frac{3 \alpha  H_0^4 (q_0-1) (16 q_0+19)+24 \pi  H_0^2 q}{27 \alpha ^2 H_0^4 (q_0-1)^2+144 \pi  \alpha  H_0^2 (q_0-1)+32 \pi ^2}\geq 0 \, ;
\end{equation}
\begin{equation}\label{16}
\rho+p\!=\!\frac{3 \alpha  H_0^4 (q_0-1) (10 q_0+13)+8 \pi  H_0^2 (q_0+1)}{27 \alpha ^2 H_0^4 (q_0-1)^2+144 \pi  \alpha  H_0^2 (q_0-1)+32 \pi ^2}\geq 0 \, ;
\end{equation}
\begin{equation}\label{17}
\rho\!=\! \frac{3 \alpha  H_0^4 (q_0-1) (7 q_0+10)+12 \pi  H_0^2}{27 \alpha ^2 H_0^4 (q_0-1)^2+144 \pi  \alpha  H_0^2 (q_0-1)+32 \pi ^2}\geq 0 \, ;
\end{equation}
\begin{equation}\label{18}
\rho-p\!=\!\frac{3 \alpha  H_0^4 (q_0-1) (4 q_0+7)-8 \pi  H_0^2 (q_0-2)}{27 \alpha ^2 H_0^4 (q_0-1)^2+144 \pi  \alpha  H_0^2 (q_0-1)+32 \pi ^2}\geq 0 \, . 
\end{equation}
\begin{figure}[t]
\centerline{\includegraphics[scale=0.25]{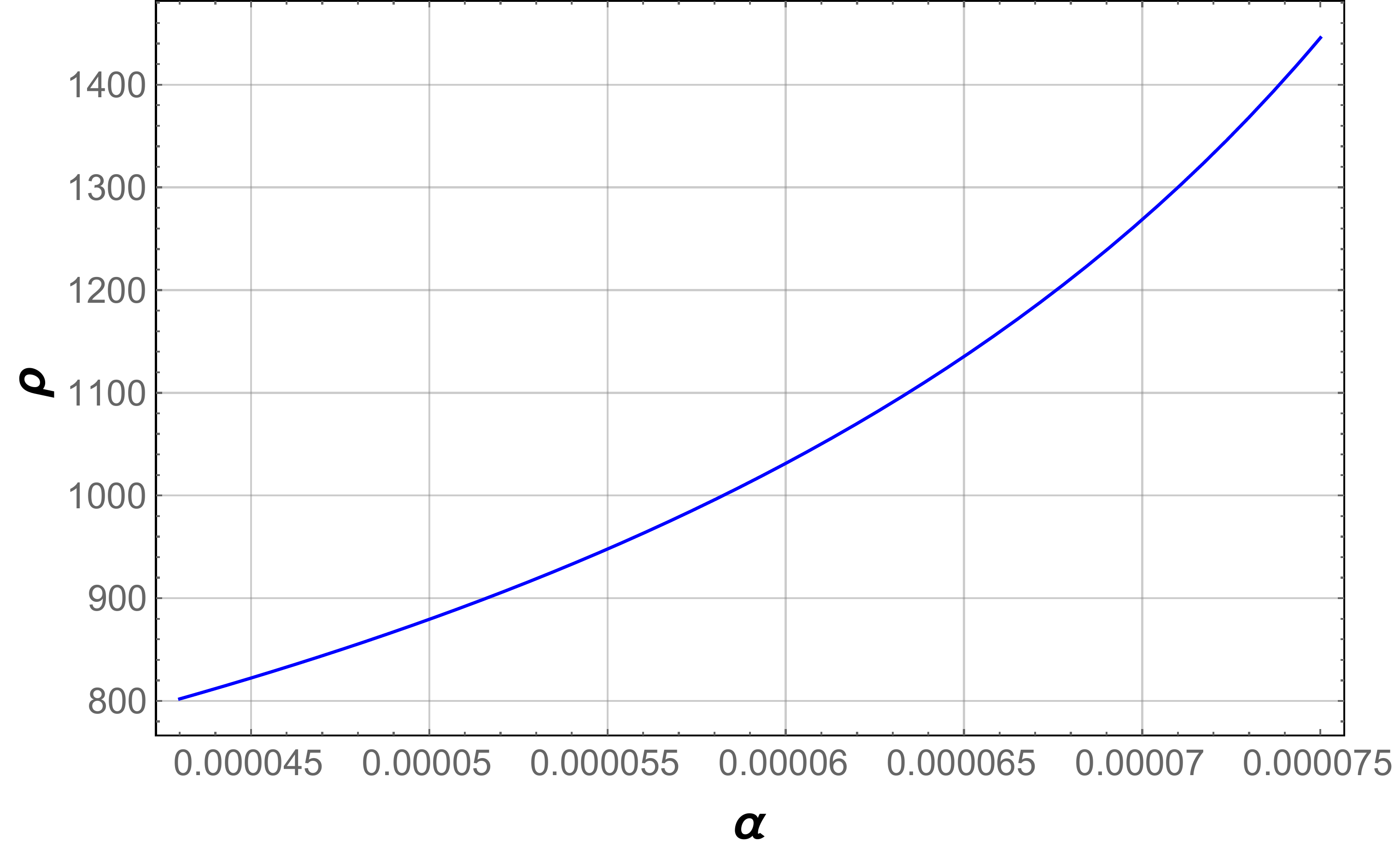}}
\caption{Energy density $\rho$ for $f(R,T)=R+\alpha RT$ derived with the present values of $H_0$ and $q_0$ parameters.}
\label{fr}
\end{figure}

\begin{figure}[t]
\centerline{\includegraphics[scale=0.25]{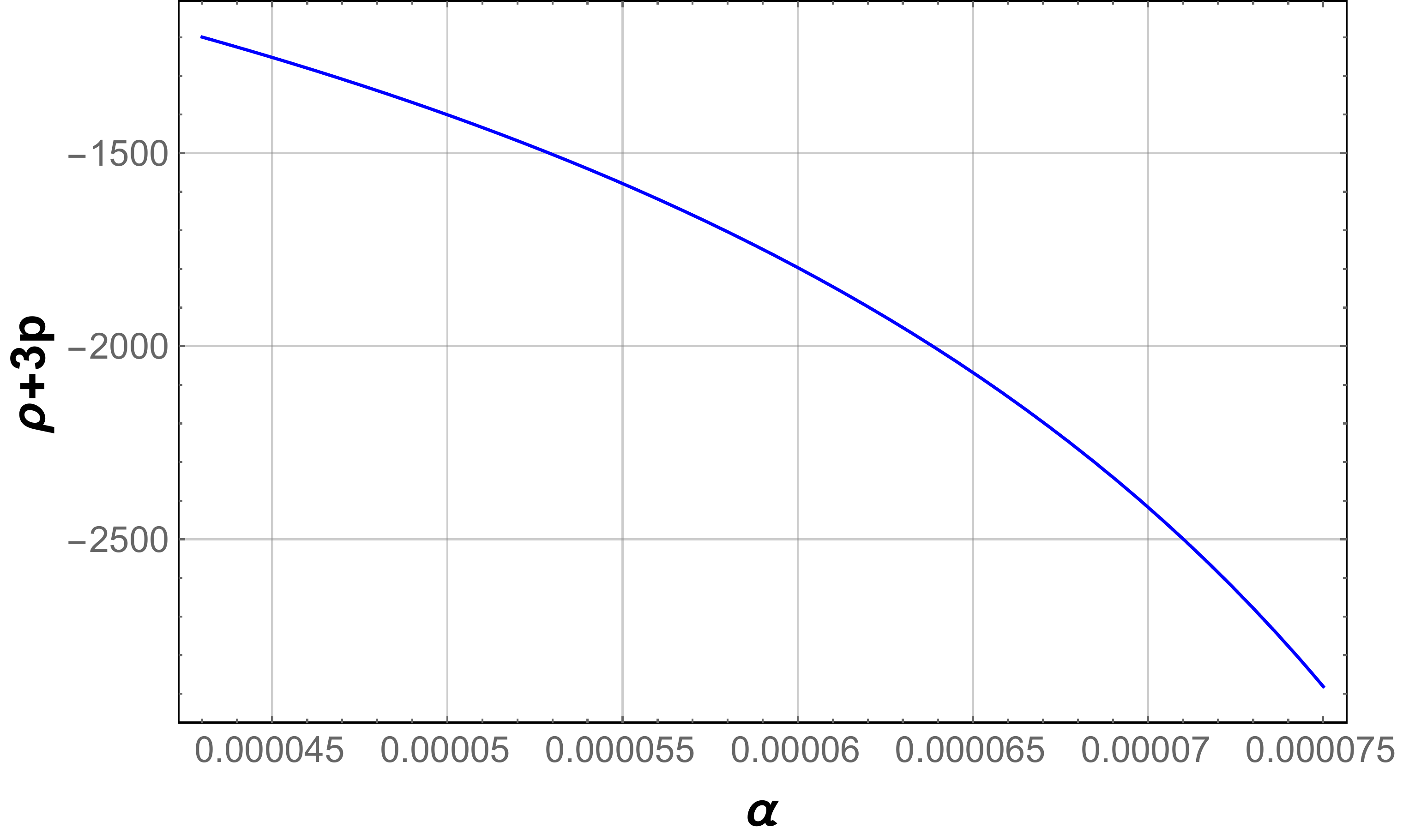}}
\caption{SEC $\rho+3p$ for $f(R,T)=R+\alpha RT$ derived with the present values of $H_0$ and $q_0$ parameters.}
\label{fsec}
\end{figure}

\begin{figure}[t]
\centerline{\includegraphics[scale=0.25]{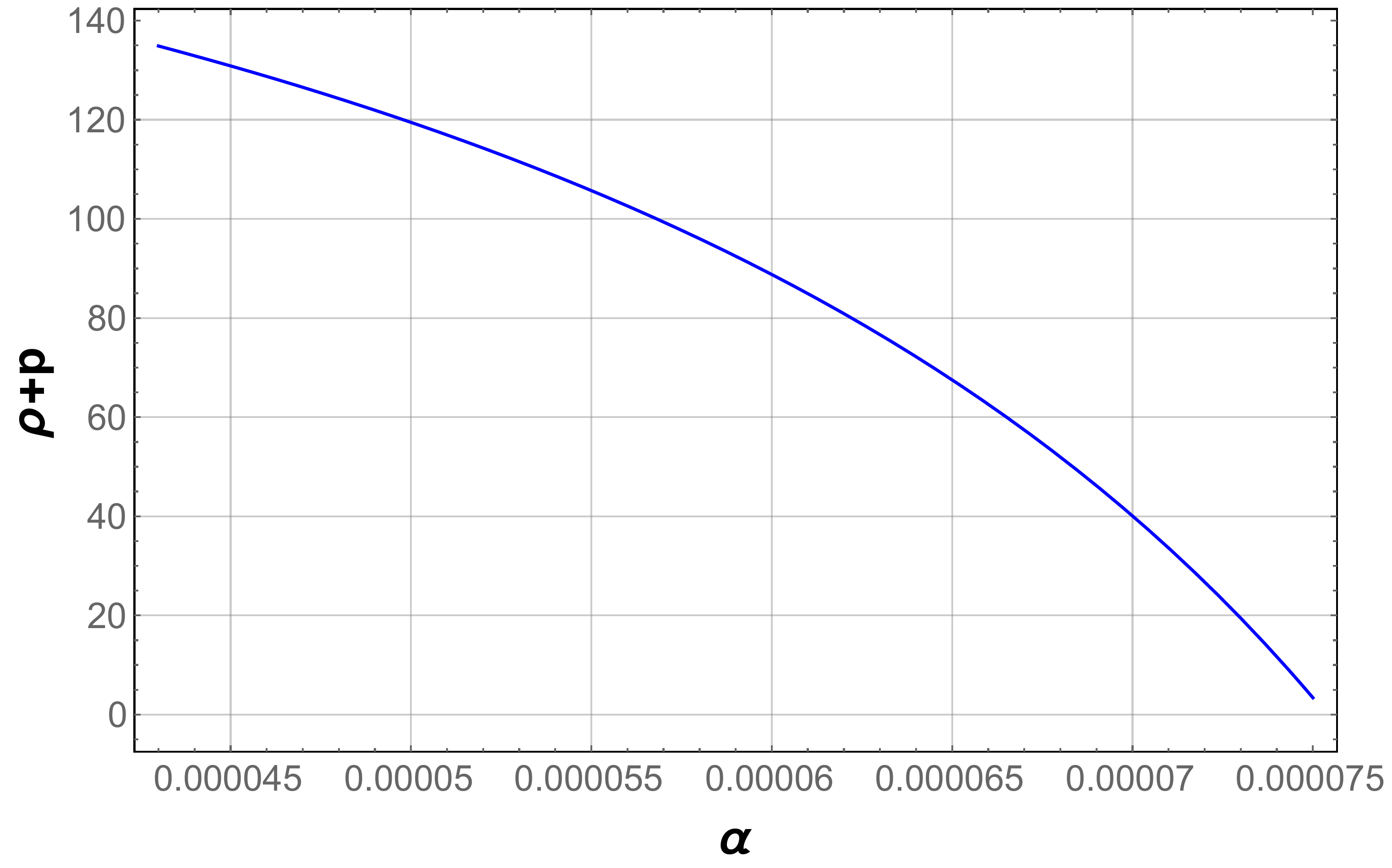}}
\caption{NEC $\rho+p$ for $f(R,T)=R+\alpha RT$ derived with the present values of $H_0$ and $q_0$ parameters.}
\label{fwec}
\end{figure}

\begin{figure}[t]
\centerline{\includegraphics[scale=0.25]{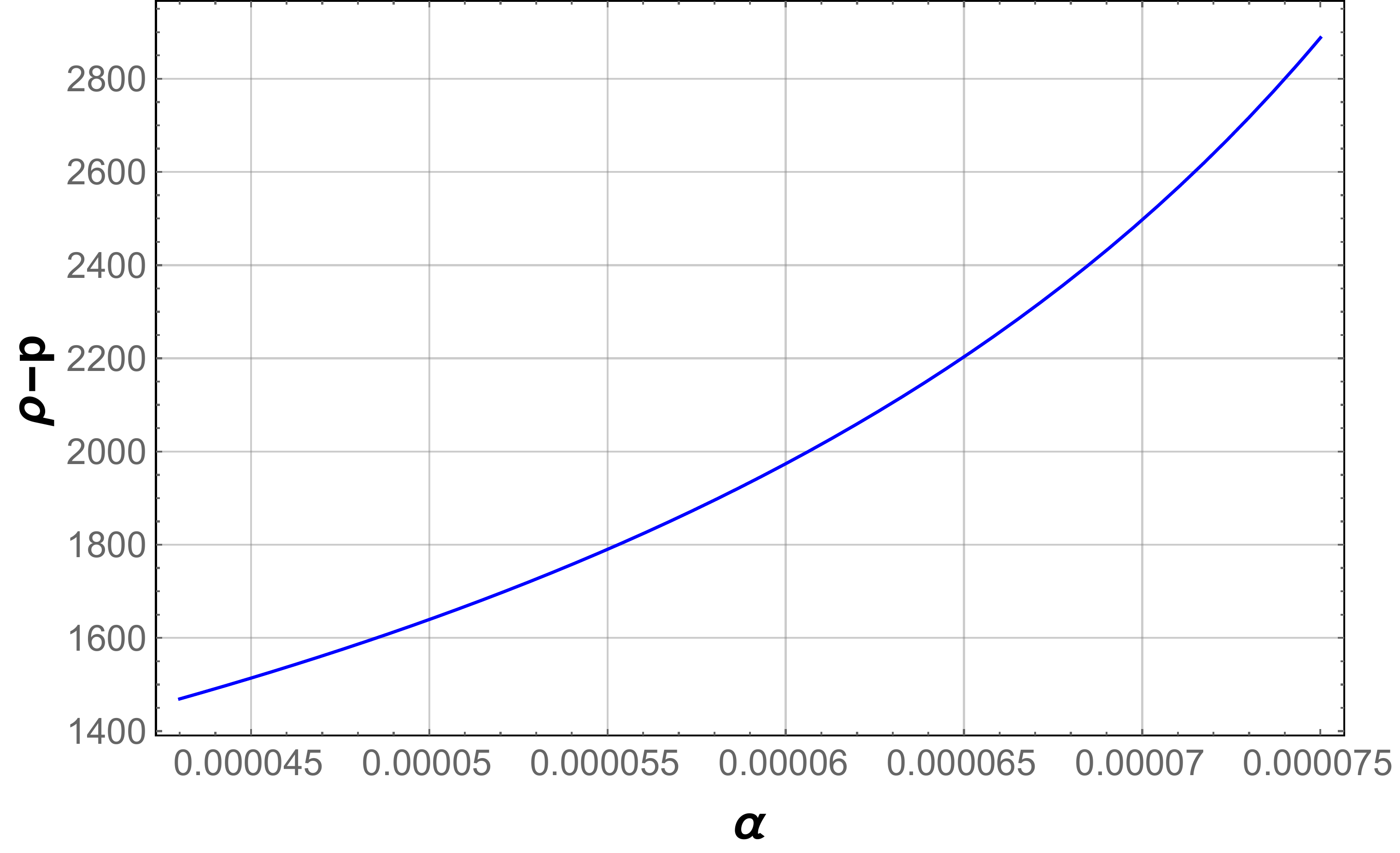}}
\caption{DEC $\rho-p$ for $f(R,T)=R+\alpha RT$ derived with the present values of $H_0$ and $q_0$ parameters.}
\label{fdec}
\end{figure}

From the above expressions \eqref{15}-\eqref{18}, one can quickly observe that the ECs depends on the model parameters $\alpha$. However, we can not choose the value of $\alpha$ arbitrarily. If we do that may cause a violation of the current accelerated scenario of the Universe. Keeping this thing in mind, we have manipulated the profiles of the ECs in Figs. \ref{fr}-\ref{fdec}.  As one can see, WEC, NEC, DEC are satisfied from Figs. \ref{fr}, \ref{fwec} \& \ref{fdec}  respectively while SEC violated with the present value of the Hubble parameter ($H_0$) and deceleration parameter ($q_0$) in Fig \ref{fsec}. Also, this is an agreement with the current scenario of the Universe.

\begin{figure}[t]
\centerline{\includegraphics[scale=0.25]{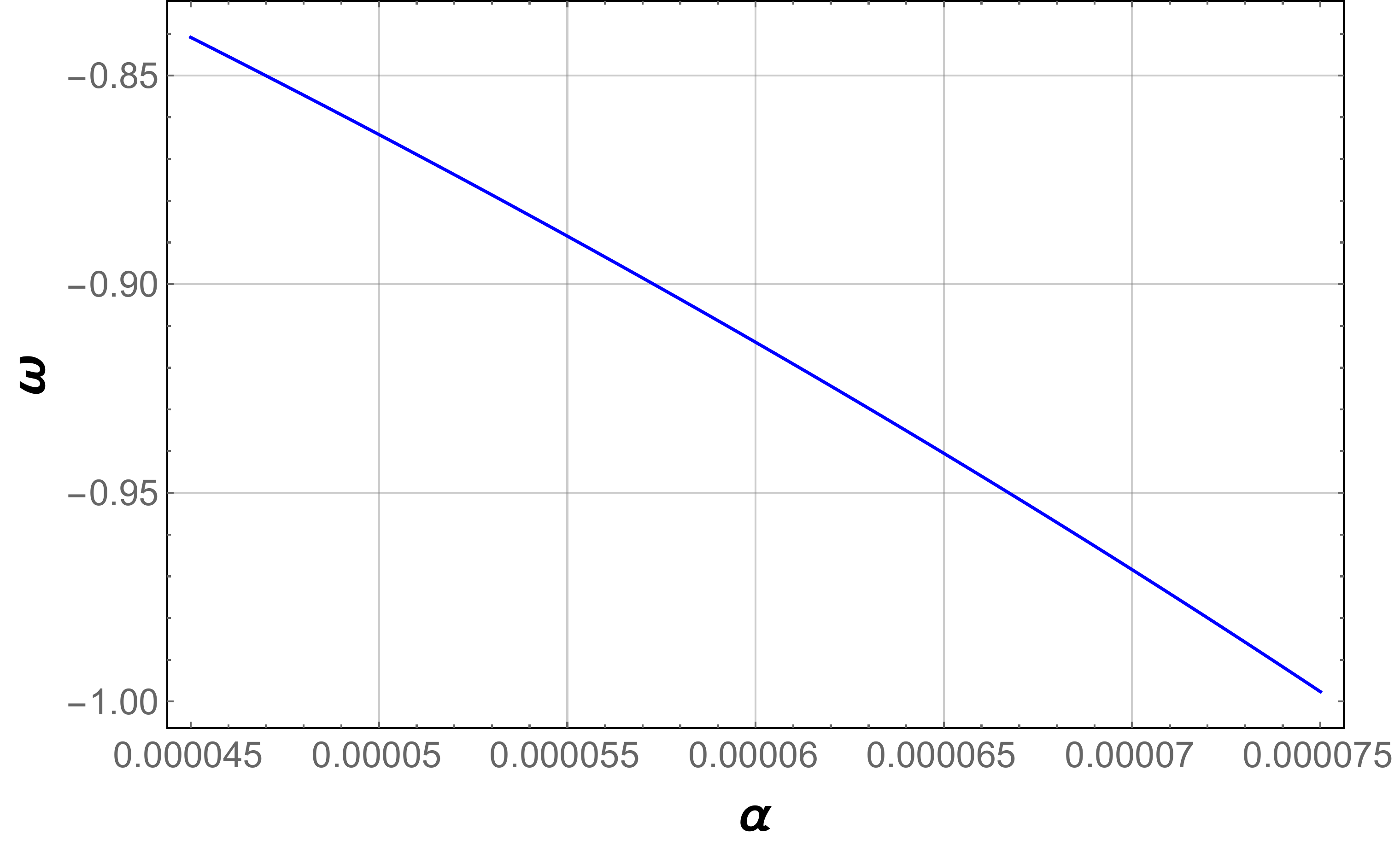}}
\caption{Equation of state parameter $\omega$ for $f(R,T)=R+\alpha RT$ derived with the present values of $H_0$ and $q_0$ parameters.}
\label{f2}
\end{figure}

The $\Lambda$CDM is a broadly accepted cosmological model that describes the observations of the Universe. Several comments and surveys, such as Planck, WAMP, and Dark Energy Survey (DES), tested it in the last two decades. A specific mapping of $f(R, T)$ mimics the $\Lambda$CDM. So, for $\alpha=0$, our model recover the $\Lambda$CDM, and the ECs for this are given by
\begin{itemize}
\item \textbf{SEC : }$6H^2 q \geq 0 \, ; $
\item \textbf{NEC : }$2H^2(1+q)\geq 0 \, ; $
\item \textbf{WEC : }$3H^2\geq 0$ and  $2H^2(1+q)\geq 0 \, ; $
\item \textbf{DEC : }$3H^2\geq 0$ and  $2H^2(1+q)\geq 0$ or, $-2H^2(-2+q)\geq 0 \, .$
\end{itemize}
By applying the present value of $H_0$ and $q_0$, one can check that the NEC, WEC, DEC satisfied whereas SEC violated. And these are the proper behavior for a standard model, which shows the accelerated expansion of the Universe. Moreover, our model is showing the same profiles for ECs, as demonstrated by $\Lambda$CDM.
Besides, the recent observation, such as the Planck collaboration and the $\Lambda$CDM, confirms the equation of state parameter $\omega $ takes its value as $\omega \simeq -1$ \citep{Planck}. Also, this is an agreement for accelerated expansion. Therefore, $\omega$ is a suitable candidate to compare our model with the $\Lambda$CDM model. The expression of $\omega$ for our model is given by
\begin{equation}
\label{19}
\omega=\frac{p}{\rho}=\frac{9 \alpha  H^2 \left(q^2-1\right)+\pi  (8 q-4)}{3 \alpha  H^2 (q-1) (7 q+10)+12 \pi } \, .
\end{equation}
In Fig. \ref{f2}, we have shown the profile of $\omega$ by taking the present value of $H_0$ and $q_0$. One can observe that $\omega$ depends entirely on the model parameter, and its value is very close to $-1$, which is in agreement with the recent observations. Moreover, our model shows the accelerated expansion as good as $\Lambda$CDM.

\section{Compatibility with Linear $f(R,T)$ Case}\label{sec5}

In the present section, we study the linear case of $f(R, T)$ model i.e., $f(R, T)= R+2 \gamma T$ \citep{Harko2011}, where $\gamma$ is a constant. Many individuals operated in $f(R, T)$ gravity in the linear case, representing energy conditions. But, by constrained values of cosmological parameters $H_{0}$ and $q_{0}$, we have attempted to research the energy conditions. The work includes the present values of $H_{0}=67.9$ and $q_{0}=-0.503$.

From \eqref{2}, one can obtain the following equations
\begin{equation}\label{22}
3H^2=(1+3\gamma)\rho-\gamma p,
\end{equation}
\begin{equation}\label{23}
2\dot{H}+3H^2=-(1+3\gamma)p+\gamma p.
\end{equation}

Using \eqref{22} and \eqref{23}, the energy density, pressure, and equation of state parameter can be obtained as,

\begin{equation}\label{24}
\rho=\frac{-2\dot{H}\gamma+3(1+2\gamma)H^2}{(1+3\gamma)^2-\gamma^2},
\end{equation}
\begin{equation}\label{25}
p=-\frac{2(1+3\gamma)\dot{H}+3(1+2\gamma)H^2}{(1+3\gamma)^2-\gamma^2},
\end{equation}
\begin{equation}\label{26}
\omega=\frac{p}{\rho}=\frac{2 H^2 (3 \gamma +1) (q+1)-3 H^2 (2 \gamma +1)}{3 H^2 (2 \gamma +1)+2 H^2 \gamma  (q+1)}.
\end{equation}
The various energy conditions reads as,

\begin{equation}\label{27}
\rho+3p=\frac{2 H_0^2 [4 \gamma +(10 \gamma +3) q_0]}{8 \gamma ^2+6 \gamma +1}\geq 0,
\end{equation}
\begin{equation}\label{28}
\rho+p=\frac{2 H_0^2 (q_0+1)}{2 \gamma +1}\geq 0,
\end{equation}
\begin{equation}\label{29}
\rho=\frac{2H_0^2(1+q_0)\gamma+3(1+2\gamma)H_0^2}{(1+3\gamma)^2-\gamma^2}\geq 0,
\end{equation}
\begin{equation}\label{30}
\rho-p=-\frac{2 H_0^2 (q_0-2)}{4 \gamma +1}\geq 0.
\end{equation}

\begin{figure}[t]
\includegraphics[scale=0.25]{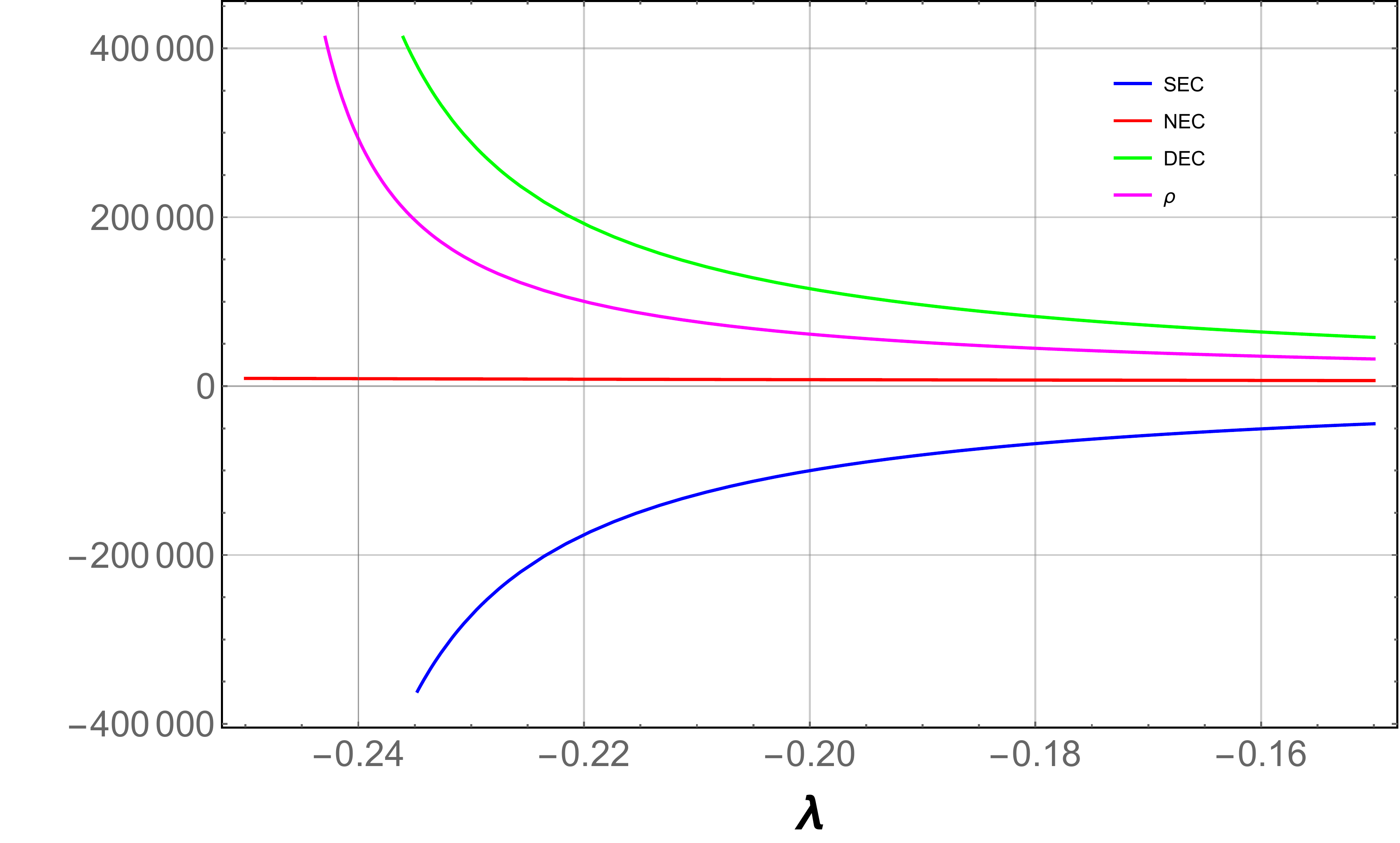}
\caption{Energy conditions for $f(R,T)=R+2\gamma T$ derived with the present values of $H_0$ and $q_0$ parameters.}
\label{f6}
\end{figure}
\begin{figure}[t]
\includegraphics[scale=0.25]{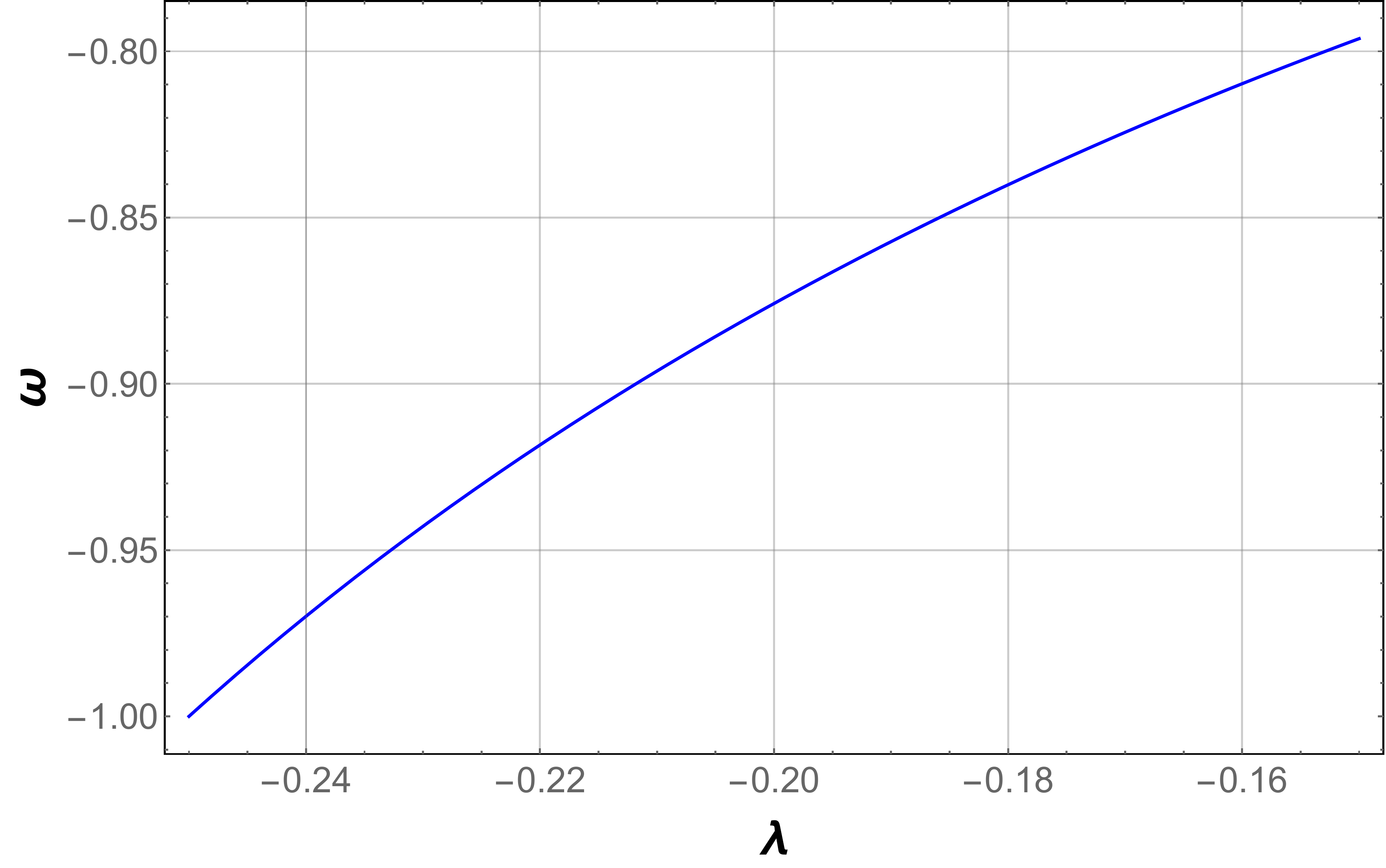}
\caption{Equation of state parameter $\omega$ for $f(R,T)=R+2\gamma T$ derived with the present values of $H_0$ and $q_0$ parameters.}
\label{f7}
\end{figure}

From the Fig. \ref{f6}, we can observe that NEC, WEC, and DEC's energy conditions fulfill their conditions for $-0.25 \leq \gamma \leq -0.15$ while SEC violates its requirement. The SEC violation represents the accelerated expansion of the universe. As we know, the EOS parameter is an acceptable candidate to compare the model with the $\Lambda$CDM, so the $\gamma$ range is selected with the EOS behavior observed.

The EOS parameter $\omega$ is approximately equal to -1 according to the Planck observations and $\Lambda$CDM. The Fig. \ref{f7} also depicts the behavior of $\omega$, taking the current values of $H_{0}$ and $q_{0}$. The values are very similar to -1, which is in line with the $\Lambda$CDM. So, $f(R, T)$'s linear and non-linear form is in good agreement with $\Lambda$CDM, resulting in an accelerated universe expansion.

\section{Conclusions} \label{sec6}

This paper built a cosmological paradigm from the simplest non-minimal matter-geometry coupling in the gravitational theory of $f(R,T)$. We have considered a well-motivated $f(R,T)$ gravity model as $f(R,T)=R+\alpha R T$, where $\alpha$ is the model parameter.

The primary motivation for such a theory of gravity is linked to its consistency with spacetime casual and geodesic structure discussed by various energy conditions. This analysis derived the null, the weak, the dominant, and strong energy conditions. In particular, the SEC plays an essential role in describing gravity's attractive or repulsive nature under the modified theory of gravity. Also, the present values of cosmological parameters $H_{0}$ and $q_{0}$ are used to check the viability of $f(R,T)$ gravity theory.

According to present values, NEC, WEC, and DEC are observed to satisfy the conditions derived from the Raychaudhuri equations, whereas SEC is violated. The energy conditions established $0.000045\leq \alpha \leq 0.000075$ constraints to describe an accelerated expansion of the Universe.We also study the linear case of $f(R, T)$ model i.e., $f(R, T)= R+2 \gamma T$, where $\gamma$ is a constant and observe that NEC, WEC, and DEC's energy conditions fulfil their conditions for $-0.25 \leq \gamma \leq -0.15$ while SEC violates its requirement. The SEC violation represents the accelerated expansion of the universe.
We also compared our energy constraints with those from the $\Lambda$CDM model. If we talk about the $\Lambda$CDM gravity, all energy conditions are satisfied except SEC. Furthermore, the equation of state parameter, derived from our model, is consistent with a current negative pressure period, showing values close to -1. This activity also confirms the explanation of $\Lambda$CDM for dark energy and recent experimental observations. So, $f(R, T)$'s linear and non-linear form is in good agreement with $\Lambda$CDM, resulting in an accelerated universe expansion.

\section*{Acknowledgements}

PKS and SA acknowledges \fundingAgency{Council of Scientific \& Industrial Research (CSIR), New Delhi, India}, for financial support to carry out the Research project No. \fundingNumber{03(1454)/19/EMR-II Dt.02/08/2019}. SM acknowledges \fundingAgency{Department of Science \& Technology (DST), Govt. of India, New Delhi}, for awarding Junior Research Fellowship file No. \fundingNumber{DST/INSPIRE Fellowship/2018/IF180676}.

%
%

%
%
%
%

\nocite{*}
\bibliography{Sahoo}%

\end{document}